\documentclass[10pt]{article}
\usepackage[utf8]{inputenc}
\usepackage[T1]{fontenc}
\usepackage{amsmath}
\usepackage{graphicx}
\usepackage[colorinlistoftodos]{todonotes}
\usepackage[colorlinks=true, allcolors=blue]{hyperref}
\usepackage{subfigure}
\usepackage{multicol}
\usepackage{caption}
\usepackage{authblk}
\usepackage{geometry}
\title{\textbf{Experimental evidence of inertial dynamics in ferromagnets}}

\author[1,$\dagger$]{Kumar Neeraj}
\author[2,$\dagger$]{Nilesh Awari}
\author[2]{Sergey Kovalev}
\author[1]{Debanjan Polley}
\author[1]{Nanna Zhou Hagström}
\author[3]{Sri Sai Phani Kanth Arekapudi}
\author[4,5]{Anna Semisalova}
\author[5]{Kilian Lenz}
\author[2]{Bertram Green}
\author[2]{Jan-Christoph Deinert}
\author[2]{Igor Ilyakov}
\author[2]{Min Chen}
\author[2]{Mohammed Bowatna}
\author[6]{Valentino Scalera}
\author[7]{Massimiliano D'Aquino}
\author[6]{Claudio Serpico}
\author[3,5]{Olav Hellwig}
\author[8]{Jean-Eric Wegrowe}
\author[9,10]{Michael Gensch}
\author[1,11]{Stefano Bonetti \footnote{e-mail: stefano.bonetti@fysik.su.se }}

\tiny{\affil[1]{Department of Physics, Stockholm University, 106 91 Stockholm, Sweden}
\affil[2]{Institute of Radiation Physics, Helmholtz-Zentrum Dresden-Rossendorf, 01328 Dresden, Germany}
\affil[3]{Institute of Physics, Chemnitz University of Technology, 09107 Chemnitz, Germany}
\affil[4]{Faculty of Physics, University of Duisburg-Essen, 47057 Duisburg, Germany}
\affil[5]{Institute of Ion Beam Physics and Materials Research, Helmholtz-Zentrum Dresden-Rossendorf, 01328 Dresden, Germany}
\affil[6]{DIETI, University of Naples Federico II, Naples, Italy}
\affil[7]{Department of Engineering, University of Naples "Parthenope", 80143 Naples, Italy}
\affil[8]{LSI, Ecole Polytechnique, CEA, CNRS, Palaiseau F-91128, France}
\affil[9]{Institute of Optics and Atomic Physics, TU Berlin, 10623 Berlin, Germany}
\affil[10]{Institute of Optical Sensor System, DLR, 12489 Berlin, Germany}
\affil[11]{Department of Molecular Sciences and Nanosystems, Ca' Foscari University of Venice, 30172 Venezia-Mestre, Italy}}

\date{}
\begin{document}

\maketitle

\begin{abstract}
The understanding of how spins move at pico- and femtosecond time scales is the goal of much of modern research in condensed matter physics, with implications for ultrafast and more energy-efficient data storage \cite{stanciu2007subpicosecond}. However, the limited comprehension of the physics behind this phenomenon has hampered the possibility of realising a commercial technology based on it. Recently, it has been suggested that inertial effects should be considered in the full description of the spin dynamics at these ultrafast time scales \cite{ciornei2011magnetization,bottcher2011atomistic, wegrowe2012magnetization,olive2012beyond,bhattacharjee2012atomistic,olive2015deviation,thonig2017magnetic,mondal2018generalisation,bastardis2018magnetization,fahnle2019comparison}, but a clear observation of such effects in ferromagnets is still lacking. Here, we report the first direct experimental evidence of inertial spin dynamics in ferromagnetic thin films in the form of a nutation of the magnetisation at a frequency of approximately 0.6 THz. 
This allows us to evince that the angular momentum relaxation time in ferromagnets is on the order of 10 ps.
\end{abstract}

The vast majority of digital information worldwide is stored in the form of tiny magnetic bits in thin film materials in the hard-disk drives installed in large-scale data-centres. The position of the north and south magnetic poles with respect to the thin film plane encodes the logical “ones” and “zeros”, which are written using strongly localised, intense magnetic fields. The dynamics of the magnetisation in the writing process is described by the Landau-Lifshitz-Gilbert (LLG) equation, which correctly models the reversal of a magnetic bit at nanosecond time-scales. Until 20 years ago, it was believed that all of the relevant physics of magnetisation dynamics was included in this equation and that optimisation of storage devices could be based solely on it.

However, the pioneering experiment of Bigot \textit{et al.} in 1996 \cite{beaurepaire1996ultrafast} revealed the occurrence of spin dynamics on the sub-picosecond scales that could not be described by the LLG equation, giving birth to the field of ultrafast magnetism. This field is currently one of the most investigated and debated topics in condensed matter physics \cite{koopmans2000ultrafast,koopmans2003experimental,koopmans2005unifying, kirilyuk2010ultrafast,battiato2010superdiffusive,mathias2012probing,radu2011transient,stamm2007femtosecond,koopmans2010explaining,boeglin2010distinguishing,carpene2008dynamics,dalla2007influence,lambert2014all,carva2013ab}, with implications for both our fundamental understanding of magnetism as well as possible applications for faster and more energy-efficient data manipulation. Recently, the LLG equation was reformulated including a physically correct inertial response, which was surprisingly missing from the original formulation, and which predicts the appearance of spin nutations, similar to the ones of a spinning top, at a frequency much higher (in the terahertz range) than the spin precession described by the conventional LLG equation, typically at gigahertz frequencies. However, the lack of intense magnetic field sources at these high frequencies has hampered the experimental observation of such nutation dynamics.

In this work, we use intense narrow-band terahertz magnetic fields at a superradiant terahertz source and the femtosecond magneto-optical Kerr effect (MOKE), to detect inertial magnetisation effects in ferromagnetic thin films. We find evidence for nutation dynamics with a characteristic frequency of the order of 1 THz which is damped on time scales of the order of 10 ps. We are able to qualitatively describe the observed magnetisation dynamics with a macrospin approximation of the inertial LLG equation and highlight implications for ultrafast magnetism and magnetic data storage.\\

\begin{figure}[t!]
\centering
\includegraphics[width=1.0\textwidth]{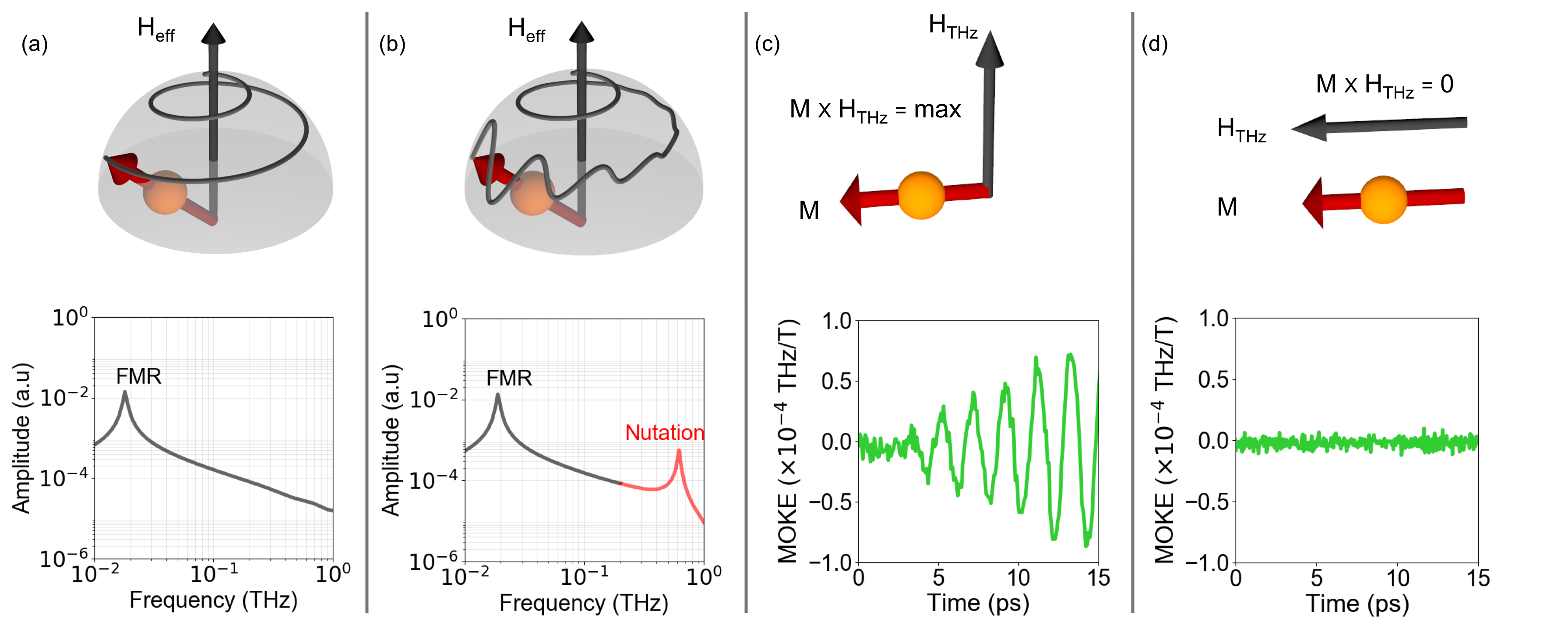}
\caption{(a) Top panel: schematic of magnetisation dynamics and relaxation around an effective magnetic field $H$ according to the standard LLG equation. Bottom panel: LLG-simulated response (i.e. the susceptibility) of a ferromagnetic system to an external ac magnetic field of varying frequency. (b) Similar to (a), but considering the inertial formulation of the LLG equation described in the text. (c) Top panel: geometrical configuration which maximises the torque of \textbf{H}$_\textrm{THz}$ on the magnetisation \textbf{M}, i.e. when they are orthogonal to each other. Bottom panel: measured response from the sample in the maximum torque configuration. (d) Similar to (c) but when \textbf{H}$_\textrm{THz}$ and \textbf{M} are parallel to each other, no torque is exerted and no signal is detectable.}
\label{f1}
\end{figure}

According to the LLG equation, the dynamics of the magnetisation \textbf{M} in a ferromagnetic sample is described as \cite{gilbert2004phenomenological}
\begin{equation} \label{eq1}
   \frac{d{\textbf{M}}}{dt} = -|\gamma| {\textbf{M}} \times \bigg({\textbf{H}_\textrm{eff}} - \frac{\alpha}{|\gamma| M_{s}}\frac{d\textbf{M}}{dt}\bigg),
\end{equation}
where $|\gamma|/2\pi \approx 28 $ GHz/T is the gyromagnetic ratio, $\textbf{H}_\textrm{eff}$ is the effective magnetic field, calculated as the variational derivative of energy with respect to the magnetisation, $M_{s}$ is the saturation magnetisation, and $\alpha$ is the so-called Gilbert damping. The first term on the right-hand-side describes the precession \textbf{M} of a ferromagnetic system around $\textbf{H}_\textrm{eff}$, while the second term is the damping term which relaxes the system to an equilibrium state where \textbf{M} and $\textbf{H}_\textrm{eff}$ are parallel and no torque is exerted on the magnetisation. Fig. \ref{f1}(a) illustrates schematically the magnetisation precession around the effective magnetic field. A resonance peak of the precession can be seen in the frequency domain and it corresponds to the so-called ferromagnetic resonance (FMR). Here, the motion of the spin system is treated as analogous to a classical spinning top. Hence, one can derive the equation of motion of spins in a magnetic field similar to that of a spinning top in a gravitational field. On a similar line, Gilbert introduced a Lagrangian for the ferromagnetic systems with an \textit{ad hoc} inertia tensor. In this mechanical approach for describing the motion of spins the two principle moments of inertia were set to zero so that the inertial terms disappear from the dynamic equation. But an inertial tensor of such kind is not physically correct as the same Gilbert noticed \cite{gilbert2004phenomenological}. Despite its crudeness, this approximation turned out to be good enough to describe the dynamics of magnetisation on time scales of 0.1 nanoseconds or longer, and the general validity of the equation at faster time scales was not questioned.

The concept of mass and inertia in macroscopic systems (domain walls) was for the first time introduced by Döring \cite{bottcher2012significance} already in 1948. In 2004 Zhu \textit{et al.} \cite{zhu2004novel} showed that the dynamics of spins in a tunnelling barrier between two superconductors has an unusual spin behaviour in contrast to a simple spin precession and they termed it as ``Josephson nutation''. Kimel \textit{et al.} \cite{kimel2009inertia} for the first time showed inertia-driven spin switching in antiferromagnetically ordered systems. According to Ref. \cite{kimel2009inertia}, a very short (femtosecond) magnetic impulse can impart enough energy to the spin system so to overcome the potential barrier and flip its orientation. But the concept of inertia in proper ferromagnetic systems arrived only in 2011 with the works of Ciornei \textit{et al.} \cite{ciornei2011magnetization}. 
A full derivation of the inertial LLG equation was later given by Wegrowe \textit{et al.} \cite{wegrowe2012magnetization}, where the three principle moments of inertia were realistically set to non-zero values, which led to the so-called inertial LLG equation.
\begin{equation} \label{eq2}
\frac{d{\textbf{M}}}{dt} = -|\gamma| {\textbf{M}} \times \left[{\textbf{H}_\textrm{eff}} - \frac{\alpha}{|\gamma| M_{s}}\left(\frac{d{\textbf{M}}}{dt} + \tau \frac{d^2 {\textbf{M}}}{dt^2}\right)\right].
\end{equation}
The last term of Eq.~(\ref{eq2}) has a second derivative term (due to angular momentum relaxation), apart from the spin precession and damping. This term comes into the picture if one considers the moment of inertia while solving the Landau-Lifshitz (LL) equation based on the classical analogue of a spinning top \cite{ciornei2011magnetization,fahnle2011generalized}. It is observed from simulations that, on time scales shorter than $\tau$, nutation oscillations are observed on top of the precession motion, as shown in Fig. \ref{f1}(b), identifying a novel ``nutation regime'' driven by inertia that has yet to be observed experimentally. On time scales longer than $\tau$ the usual LLG equation is recovered. The possible large separation between the time scales of the two regimes would then explain the success of the standard LLG equation in correctly describing magnetisation dynamics for times larger than $\tau$. The determination of the value of $\tau$ is however an open problem which needs to be addressed experimentally, as different theoretical works indicate values ranging from a few femtoseconds to hundreds of picoseconds.\\

In order to attack this problem, one needs to be able to perform magnetic field spectroscopy in the terahertz range, a task which was technically unfeasible until very recently, when intense terahertz sources have started to become available. In the past few years, broadband terahertz radiation generated with table-top laser sources \cite{hoffmann2011intense} has been exploited to study magnetisation dynamics in different classes of materials \cite{kampfrath2013resonant,vicario2013off, bonetti2016thz,hafez2018extremely,noe2018coherent,polley2018thz}. In addition to demagnetisation effects similar to those observed with near-infrared radiation, the terahertz magnetic field $\mathbf{H}_\textrm{THz}$ of such intense radiation can exert a Zeeman torque $\mathbf{M} \times \mathbf{H}_\textrm{THz}$, leading to a coherent precessional motion of the magnetisation vector \textbf{M} which lasts until the THz pulse has left the material. Those table-top sources, while having the necessary terahertz bandwidth to perform the proposed magnetic field spectroscopy, are not spectrally dense, and the detection of a nutation resonance driven by inertia has not been reported yet.

\begin{figure}[t!]
\centering
\includegraphics[width=0.66\textwidth]{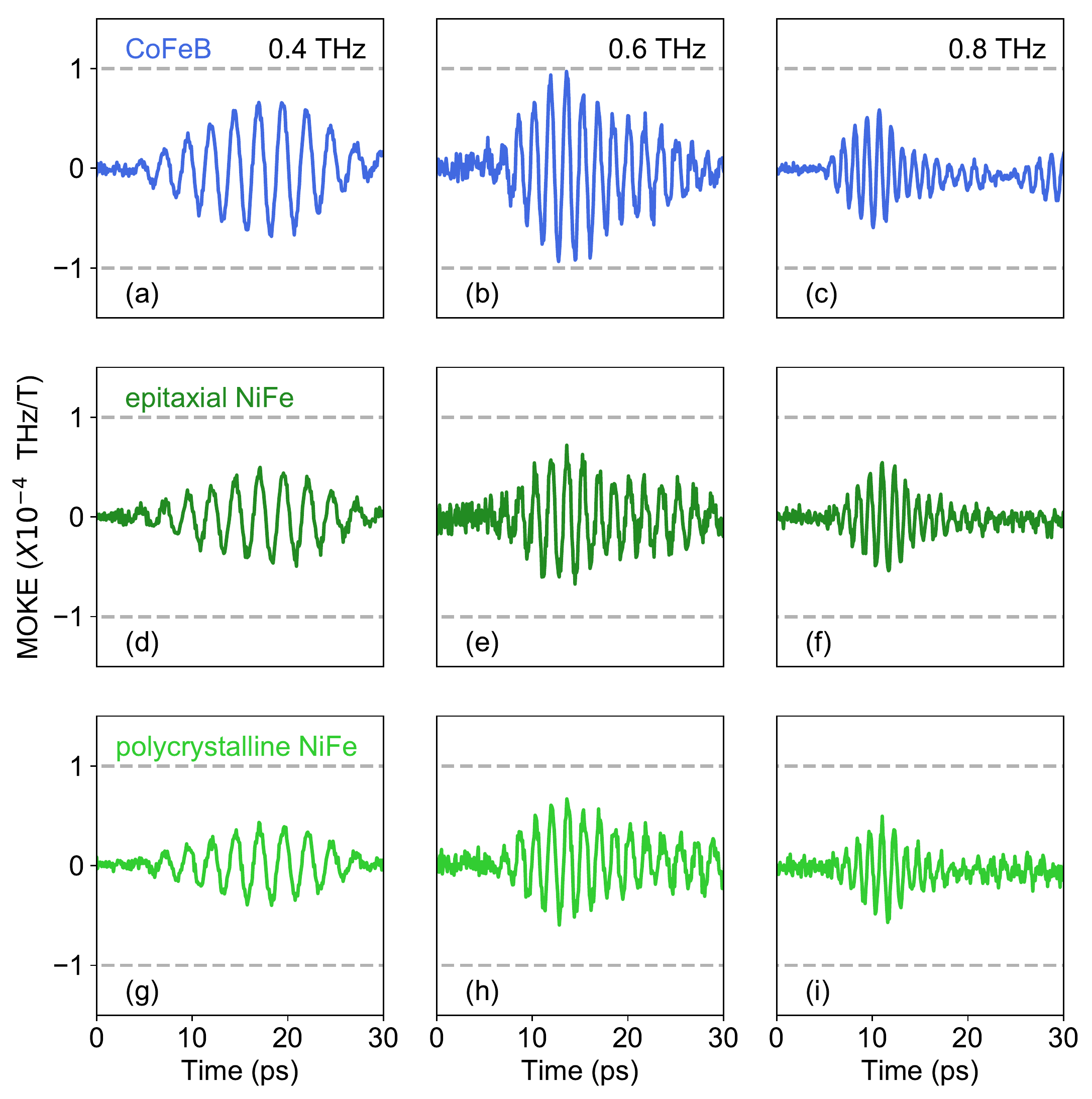}
\caption{Time-resolved magneto-optical Kerr (MOKE) response of the magnetisation to narrowband terahertz fields centered around 0.4, 0.6 and 0.8 THz for (a)-(c) an amorphous CoFeB film on silicon, (d)-(f) an epitaxial Ni$_{81}$Fe$_{19}$ (permalloy) film grown on MgO (100) substrate and (g)-(i) of a polycrystalline Ni$_{81}$Fe$_{19}$ deposited on MgO (111) substrate. The data is normalised according to Eq.~(\ref{eq3}) and related considerations in the main text.}
\label{f2}
\end{figure}

\begin{figure}[t!]
\centering
\includegraphics[width=0.66\textwidth]{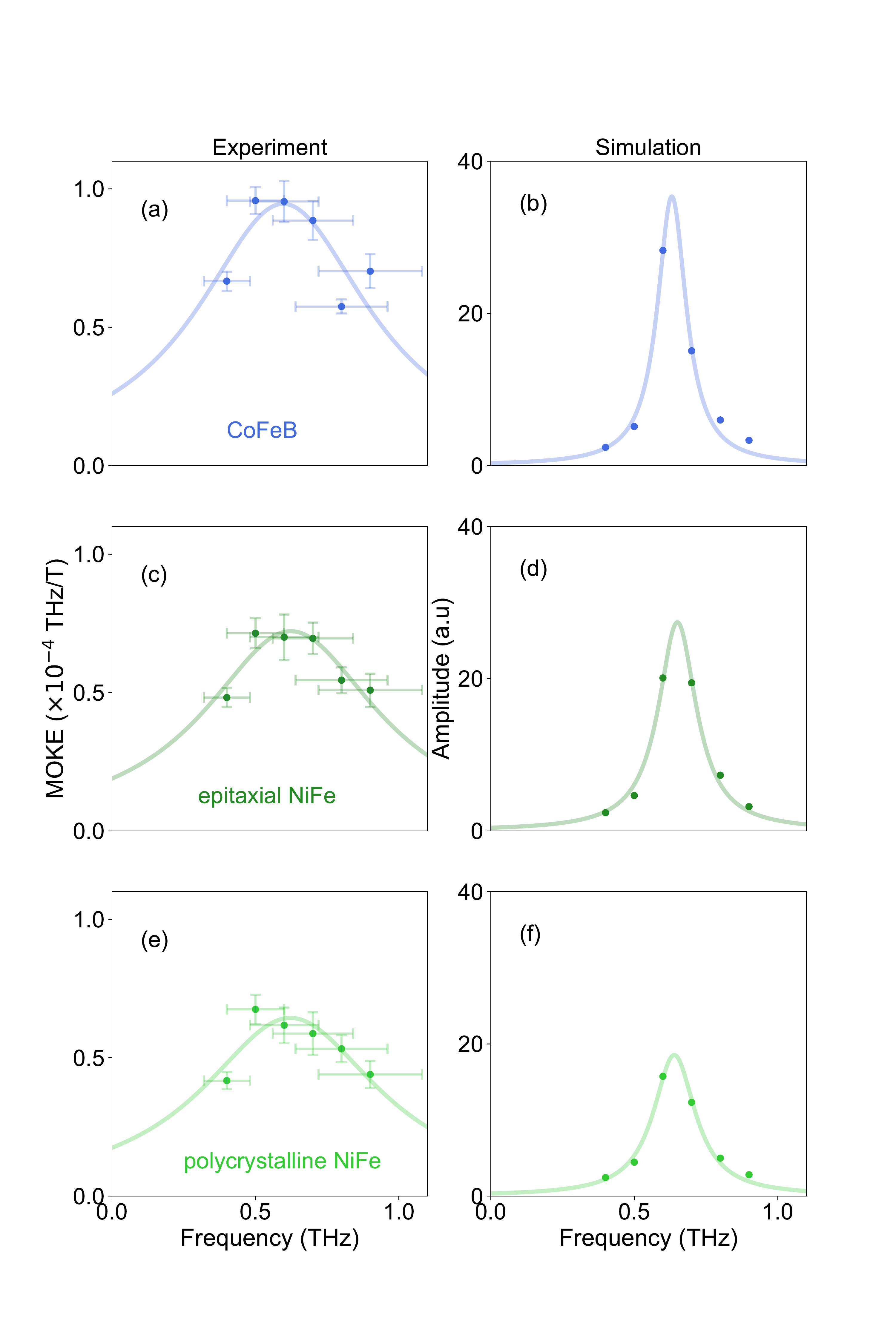}
\caption{Symbols: (a), (c), (e) experimentally measured maximum MOKE amplitude normalized according to Eq.~(\ref{eq3}) for CoFeB, epitaxial NiFe and, respectively, polycrystalline NiFe; (b), (d), (f) calculated maximum magnetisation response amplitude solving the inertial LLG equation. Solid lines: Lorentzian fit to the data points.}
\label{f4}
\end{figure}

However, intense and tunable narrow-band terahertz magnetic fields can now be generated at superradiant electron sources such as the TELBE facility in Dresden, Germany \cite{kovalev2017probing,kovalev2018selective}. Here, we used the radiation generated at TELBE to drive the magnetisation of thin film ferromagnets with a strong oscillating terahertz magnetic field. The basic idea is to perform a forced oscillator experiment as a function of the frequency of the terahertz magnetic field $\mathbf{H}_\textrm{THz}$, detecting amplitude and phase of the response with the femtosecond MOKE, in the attempt to observe the signature of a resonance. The response of the magnetisation is maximised when $\mathbf{H}_\textrm{THz}$ and the static magnetisation (which is controlled with an external magnetic field) are orthogonal to each other. This is illustrated schematically in the top panels of Fig. \ref{f1}(c) and \ref{f1}(d), with the corresponding experimental measurement in the bottom panels, where we show the detected polar MOKE signal. The polar MOKE configuration is the optimal geometry since the torque acts out of the film plane when both $\mathbf{H}_\textrm{THz}$ and $\mathbf{M}$ are in plane. Further details of the experimental setup are given in the Methods section. However, we comment here on the normalisation procedure, which is a key aspect when measuring absolute amplitudes. With the realistic assumptions that the first term on the right-hand side of Eq.~(\ref{eq1}) and Eq.~(\ref{eq2}) is much larger than the other terms, and that the magnetisation responds linearly to the field, one can read the response of the magnetisation as the integral of the terahertz magnetic field. Approximating the terahertz magnetic field as a sinusoidal excitation with angular frequency $\omega$, in the case of orthogonal $\mathbf{M}$ and $\mathbf{H}_\textrm{THz}$, the temporal response of the normalised magnetisation $m(t)=M(t)/M_s$ ignoring resonance effects can be simplified as
\begin{equation} \label{eq3}
m(t) = |\gamma| \int H_{\rm THz} \sin \omega t \, dt = |\gamma|\frac{H_{\rm THz}}{\omega}\cos\omega t,
\end{equation}
with the magnetisation starting to move out of the plane of the film when $\mathbf{M}$ and $\mathbf{H}$ are in the film plane. Hence, in order to estimate the proper response of the magnetic films to the terahertz magnetic field, the data presented in Fig. \ref{f2} and Fig. \ref{f4} has been normalised by the magnitude $H_{\rm THz}$ and by the frequency $\omega$ of the THz pulse.

We investigated three different thin film samples, all with easy-plane magnetisation: amorphous CoFeB grown on Si/SiO$_2$ substrate, whereas epitaxial and polycrystalline permalloy were grown on single crystal MgO (100) and (111) substrates respectively. Those samples have different Gilbert damping parameter $\alpha$, and saturation magnetisation M$_s$. Both parameters modulate the magnitude of the additional inertial term introduced in Eq.~(\ref{eq2}).

Figure \ref{f2} shows the amplitude of the femtosecond MOKE response of the three ferromagnetic thin film samples after excitation with narrowband terahertz pulses with a centre frequency of 0.4, 0.6 and 0.8 THz. The characterisation of the narrowband terahertz pulses, performed with electro-optical sampling, is described in the Methods section. The terahertz magnetic field was applied orthogonal to the equilibrium magnetisation direction as discussed above, so to maximise the torque. In all cases, a clear coherent response of the magnetisation to the narrowband terahertz field is observed. After the normalisation procedure discussed above, the overall response seems to be slightly larger for the CoFeB samples, likely related to the larger magneto-optical response of that sample as compared to permalloy. Hence, to infer real magnetic effects (as compared to magneto-optical ones), absolute amplitudes should be compared only within the same sample at different frequencies, and not between different samples. With this consideration, one can observe an evident change in the amplitude of the response within each sample when the frequency is changed, with the largest effect observed at 0.6 THz for all three cases.

\begin{figure}[t!]
\centering
\includegraphics[width=0.66\textwidth]{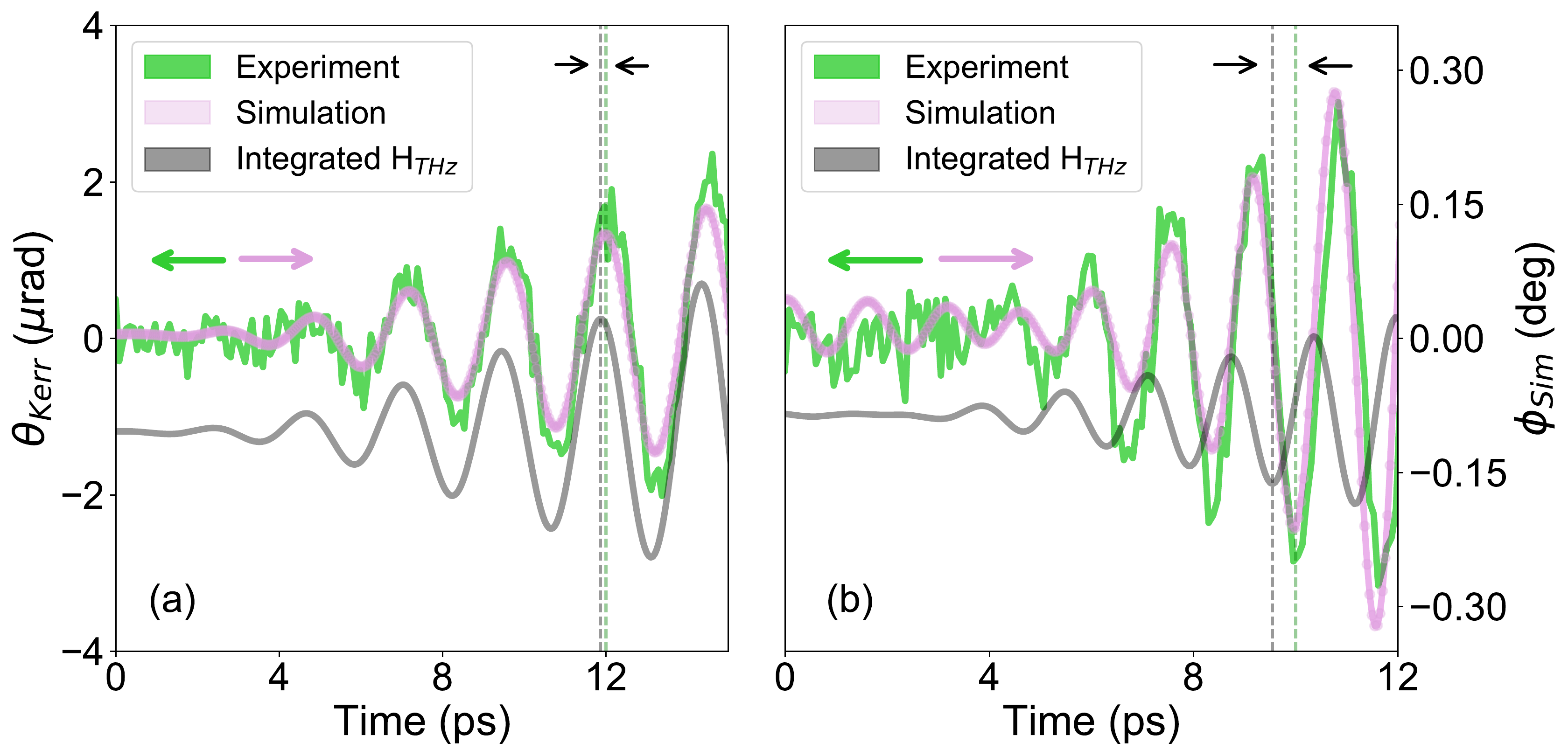}
\caption{Comparison of the phase resolved response at (a) 0.4 THz and (b) 0.6 THz center frequency of the terahertz magnetic field pulse for the polycrystalline permalloy film. Green curves: experimentally measured magneto-optical Kerr rotations, without the scaling described by Eq.~(\ref{eq3}). The absolute value of rotation for both frequencies are plotted on the left vertical axis. Pink curves: simulated response using the inertial LLG equation with $\tau=11.3$ ps and using the experimentally measured $H_{\rm THz}$ field amplitude. The right vertical axis is the simulated nutation angle. Grey curves: time integral of the experimental terahertz magnetic field $H_{\rm THz}$.}
\label{f3}
\end{figure}

The observed response of the magnetisation confirms the analogy with a forced oscillator, where the terahertz magnetic field acts as the driving periodic force to which the magnetisation responds, in the linear regime, by integrating it. The modulation of the amplitude of the response suggests already the presence of an underlying resonance at approximately 0.6 THz superimposed to the purely off-resonant, forced response of lower amplitude. To further validate this point, in Fig.~\ref{f3} we analysed the relative phase shift between the integral of the driving force (the terahertz magnetic field, which is reconstructed independently via experimental electro-optical sampling) and the experimental MOKE signal far away from the maximum response (0.4 THz) and at the maximum response (0.6 THz). We show here data for one of the samples, but similar trends are observed for the other two (see Supplementary Information). The data shows that at 0.4 THz the magnetisation precession is in phase with the driving field, and this is also reproduced by simulations solving the inertial LLG equation \cite{olive2012beyond, Li2015inertial} including the experimentally measured terahertz magnetic field as the driving force. (See Methods section for details.) At 0.6 THz, on the other hand, magnetisation precession and driving field are approximately 90 degrees out of phase, which is also reproduced by the simulations. This fact reinforces the statement that an underlying resonance is present in the system, at a frequency two orders of magnitude higher than any known ferromagnetic resonance, and which is determined by one single fitting parameter $\tau$, i.e. the angular momentum relaxation term in Eq.~(\ref{eq2}).

In order to estimate the value of $\tau$ from experiments, we plot in Fig.~\ref{f4}(a) the amplitude of the measured response at six different frequencies. A fit with a Lorentzian curve was used to return the centre frequency $\omega_n$. According to Ref.~\cite{olive2015deviation}, the nutation frequency is
\begin{equation}
\omega_n = \dfrac{\sqrt{1+\alpha \tau |\gamma| H}}{\alpha |\gamma|}\approx1/\alpha \tau.
\label{eq:wn}
\end{equation}
We measured the Gilbert damping $\alpha$ independently with ferromagnetic resonance spectroscopy in all three samples and extracted the corresponding $\tau$, as summarised in Table~\ref{t1}.  Numerical calculations that solve the inertial LLG equation with these values of $\tau$ are shown in Fig.~\ref{f4}(b), and reproduce the main features of the experimental data.

\begin{table}[]
    \centering
    \begin{tabular}{|c|c|c|c|c|}
    \hline
         Sample  & Center frequency $\omega_n/2\pi$ (THz) & FWHM (THz) & $\alpha$& $\tau = 1/\alpha\omega_n$ (ps)\\
         \hline
         CoFeB & $0.59\pm0.13$ &0.52& 0.0044& $60.5\pm13.3$\\
         epitaxial NiFe & $0.62 \pm 0.12$ & 0.57 & $0.0058$ & $44.0\pm8.5$ \\
         polycrystalline NiFe  & $0.61 \pm 0.12$& 0.58 & $0.0230$ & $11.3 \pm 2.2$\\
    \hline
    \end{tabular}
    \caption{The center frequency $\omega_n$ and the full-width half maximum (FWHM) were extracted parameters from the Lorentzian fit of the experimental data plotted in Fig. \ref{f4} for all three samples. The Gilbert damping $\alpha$ was measured independently. The angular momentum relaxation time $\tau$ is calculated using the approximation of Eq.(\ref{eq:wn}).}
    \label{t1}
\end{table}

We now make several considerations looking at Fig.~\ref{f4} and Table~\ref{t1}. The first observation is that the peak centre frequency moves only slightly between the different samples. This may be due to the fact that the saturation magnetisation $M_s$ and the Gilbert damping $\alpha$ are reasonably similar, i.e. within the same order of magnitude, in these three samples. The nutation frequency $\omega_n$ is also very weakly dependent on the applied magnetic field. We applied an in-plane field of up to 100 mT, but did not observe any measurable effect on the MOKE response. Solving the inertial LLG equation, we estimated that we would need a 10 T applied field to shift the peak by approximately 200 GHz, which should be detectable even given the large width of the resonance. We did not have the capabilities to apply such a large field in the experiments presented here, but we aim at investigating the dynamics under such magnetic field conditions in the future. 
An alternative explanation for the presence of high-frequency magnetic excitations could be the presence of exchange-dominated standing spin-wave modes across the film thickness, similar to what was observed in Ref. \cite{razdolski2017nanoscale}. However, we argue that those modes are not relevant here, simply because they cannot be excited in our experiment. In fact, we directly drive the film with electromagnetic radiation whose $k$-vector is orders of magnitude larger than the film thickness.


A second important observation is that the width of the experimental peak is much larger than what simulations predict; this mismatch can be attributed to the fact that the inertial motion of the spins might be affected by the microscopic details of the material. For a proper comparison, micromagnetic simulations based on the inertial LLG equation would need to be performed. The inertial dynamics has been reported to remain the same if one considers the dynamics of a macroscopic magnetic volume element \cite{mondal2018generalisation}. In addition, our simulations were performed at a temperature $T=0$ K, and some linewidth broadening should be expected when finite temperature is considered.

Finally, we comment on the experimentally extracted values of $\tau$ of the order of 10 ps. These are 1 to 2 orders of magnitude larger than the ones obtained with indirect spectroscopy experiments observing the stiffening of the ferromagnetic resonance, but with no phase information \cite{Li2015inertial}. This is a discrepancy that we cannot reconcile at the moment, but we envision that future experiments with terahertz sources able to explore the 10 THz region and above will be able to tell on whether other resonance peaks can be observed at even higher frequencies. At present, our data clearly demonstrate the existence of a resonance in the terahertz regime where the most plausible theoretical explanation is the one of nutation dynamics being excited in the magnetic system.

In conclusion, we used narrowband terahertz magnetic fields to drive magnetisation dynamics in thin ferromagnetic films, which we probed with the femtosecond magneto-optical Kerr effect. By analysing both amplitude and phase of the response, we detect the appearance of a broad resonance at approximately 0.6 THz, which we ascribe to the presence of a nutation spin resonance excited by the terahertz magnetic field. Our experimental observations are in good agreement with numerical simulations performed with the inertial version of the LLG equation using the angular momentum relaxation time $\tau$ extracted from the experimental data. We anticipate that our results will allow for a better understanding of the fundamental mechanisms of ultrafast demagnetisation and reversal, with implications for the realisation of faster and more efficient magnetic data storage.

\section*{Acknowledgements}
We thank Jürgen Lindner (HZDR, Dresden) for helpful discussion. The research leading to this result has been supported by the project CALIPSOplus under Grant Agreement 730872 from the EU Framework Programme for Research and Innovation HORIZON 2020. The authors acknowledge the ELBE team for operating the ELBE facility. S.K., B.G., and M.G. acknowledge support from the European Cluster of Advanced Laser Light Sources (EUCALL) project, which has received funding from the European Union's Horizon 2020 research and innovation programme under Grant Agreement No 654220. N.A., I.I., M.G. and S.K. acknowledge support from the European Commission’s Horizon 2020 research and innovation programme,under Grant Agreement No DLV-737038(TRANSPIRE). K.N., D.P., N.Z.H. and S.B. acknowledge support from the European Research Council, Starting Grant 715452 MAGNETIC-SPEED-LIMIT.

\section*{Author Contributions}
S.B. designed the experiment, S.B. and M.G. coordinated the project; K.N., N.A., S.K., D.P., N.Z.H., B.G., J.C.D., I.I., M.C., M.B., M.G. and S.B. performed the measurements at TELBE; K.N., N.A. and S.B. performed the data analysis; K.N., V.S., M.A. and C.S. performed the inertial LLG simulations. S.S.P.K.A., O.H, A.S., K.L. fabricated and characterised the samples; K.N. and S.B. coordinated the work on the paper with contributions from N.A., S.K., S.S.P.K.A., O.H., A.S., K.L., J.E.W., M.G. and discussions with all authors.

\section*{Corresponding authors}
Correspondence to S. Bonetti.

\section*{Methods} \label{methods}
\subsection*{Sample details}

Epitaxial permalloy (Ni$_{81}$Fe$_{19}$ at. $\%$) films investigated in the study are 15 nm thick and prepared by magnetron sputter deposition using Ar sputter gas at $P_{Ar}= 3.5 \times 10^{-3}$ mbar. At first, the single crystal MgO (100) substrates were pre-annealed at 873 K for 3 hours to remove typical inorganic Mg(OH) layers from the surface, then the substrate was cooled down to 550 K for DC - magnetron sputtering. The epitaxial relation between the substrate and the films (MgO(100)[100]$_{fcc}$||Ni$_{81}$Fe$_{19}$(100)[100]$_{fcc}$, c/a ratio of 0.99) were confirmed by XRD ($\phi$-scans along (111) orientation of the MgO crystal and the NiFe film). 
After the same substrate pre-annealing step as used above (873 K for 3 hours), 15 nm thick polycrystalline Ni$_{81}$Fe$_{19}$ films were deposited at room temperature on a MgO (111) substrate. In both cases (epitaxial and polycrystalline Py), an Al cap layer of 1.5 nm was deposited at room temperature to protect the magnetic films from oxidation. The atomic stoichiometry of the Ni$_{81}$Fe$_{19}$ alloy thin-films were determined by Rutherford Backscattering Spectrometry (RBS) within the measurement accuracy of $\pm1$ at.\%.  

Amorphous 10 nm thick Co$_{40}$Fe$_{40}$B$_{20}$ films were sputter-deposited at room temperature from a stoichiometric target on a Si/SiO$_{2}$(100 nm)/Al$_{2}$O$_{3}$(10 nm) substrate, followed by an Al (1.5 nm) cap layer. Commercial 2-inch stoichiometric alloy targets of Ni$_{81}$Fe$_{19}$, Co$_{40}$Fe$_{40}$B$_{20}$, (4N material purity) were produced by liquid metallurgy. The deposition rates were pre-calibrated using X-ray reflectivity measurements (XRR) and during the sputtering process, the thicknesses of the individual layers were monitored via a quartz crystal. The saturation magnetisation $M_s$ was determined from a ~3 mm $\times$ 3 mm sample pieces using SQUID-VSM, and we found the values $M_s = 756$ kA/m$^3$ ($\mu_0M_s=0.95$ T) for epitaxial permalloy, $M_s = 738$ kA/m$^3$ ($\mu_0M_s=0.93$ T) for polycrystalline permalloy, and $M_s = 1290$ kA/m$^3$  ($\mu_0M_s=1.62$ T) for amorphous CoFeB.

\subsection*{Experimental setup}
The superradiant THz pulse source, TELBE, is capable of delivering multicycle electromagnetic pulses with a maximum repetition rate of 100 kHz, which can be tuned from 0.1 to 1.3 THz. The generated THz pulses are linearly polarized and consists of around 8 cycles with a spectral bandwidth of 20 \% \cite{kovalev2018selective}. The THz electric field was measured in time domain by free-space electro-optic sampling in a 100 $\mu$m thick ZnTe crystal. A 100 fs duration laser pulse with a central wavelength of 805 nm, from a commercial Ti:sapphire laser system is used as a probe. The laser is synchronised and timed with the TELBE source \cite{green2016high}. There is an uncertainty of typically 30 fs (FWHM) in the timing between the probe laser and the THz pulse generated from the undulator \cite{kovalev2018selective}. 



In the experiment here described, the THz radiation from TELBE is focused onto the sample using gold parabolic mirrors. In our measurement scheme, the samples were magnetised in plane using a static bias magnetic field of 0.35 T. The coherent response of the sample magnetisation to the THz field was probed by measurements of the magneto-optical Kerr effect. The dependence of the induced dynamics on the THz frequency of the pump pulses was studied by tuning the electromagnetic undulator \cite{kovalev2018selective}. A femtosecond laser is used to probe the dynamics of the polar component of the sample magnetisation using MOKE. 





\subsection*{MOKE data normalization}
The normalized MOKE curves in Fig. \ref{f2} obtained from the reflectivity data were calculated as: 
\begin{equation}
    \centering
    MOKE = \frac{\Delta R}{2R} \times \frac{\omega /2\pi}{H_\textrm{THz}},
\end{equation}
where $\Delta R$ is the transient reflectivity change due to the THz pump, $\omega$ is the center frequency of the terahertz pump field and $H_\textrm{THz}$ its maximum amplitude. Error bars in the resonance amplitude are calculated by taking half of the standard deviation of the measurement before time zero. The error bars on the frequency axis are calculated considering the fact that the bandwidth of the TELBE source is approximately 20\%.  
\bibliographystyle{naturemag}
\bibliography{Reference}
\end{document}